\pdfoutput=1

\documentclass[3p,times,twocolumn]{elsarticle} 

\usepackage{epsfig}
\usepackage{graphicx}
\usepackage{amsmath}

\usepackage{amssymb}

\usepackage[figuresright]{rotating}

\usepackage{hyperref}

\usepackage{amssymb}
\usepackage{subfigure}
\def\bea {\begin{eqnarray}}
\def\eea {\end{eqnarray}}
\def\ra {\rightarrow}
\def\be {\begin{equation}}
\def\ee {\end{equation}}
\def \beq{\begin{equation}}
\def \eeq{\end{equation}}
\def \beqa{\begin{eqnarray}}
\def \eeqa{\end{eqnarray}}
\def \la{\langle}
\def \ra{\rangle}
\def \l{\left(}
\def \r{\right)}
\def \l{\left(}
\def \r{\right)}

\newcommand{\pT}{$p_{\rm{T}}$}
\newcommand{\kT}{$k_{\rm{T}}$}

\newcommand{\muB}{$\mu_{\rm B}$}

\newcommand{\sNN}{$\sqrt{s_{\rm NN}}$}

\usepackage{lineno}

\begin{document}
\begin{frontmatter}

\title{
Isothermal compressibility of hadronic matter formed in \\relativistic nuclear collisions}

\author[a]{Maitreyee Mukherjee\corref{corr1}}
\cortext[corr1]{maitreyee.mukherjee@cern.ch; Presently at CCNU, Wuhan, 430079, PR China}
\author[b]{Sumit Basu\corref{corr2}}
\cortext[corr2]{sumit.basu@cern.ch}
\author[c]{Arghya Chatterjee}
\author[d]{Sandeep Chatterjee}
\author[c]{Souvik Priyam Adhya}
\author[c]{Sanchari Thakur}
\author[c,e]{Tapan K. Nayak}
\address[a]{Bose Institute, Department of Physics and CAPSS, Kolkata-700091, India} 
\address[b]{Department of Physics and Astronomy, Wayne State University, Detroit, MI 48201, USA}
\address[c]{Variable Energy Cyclotron Centre, HBNI, Kolkata-700064, India}
\address[d]{AGH University of Science and Technology, al. Mickiewicza 30, 30-059 Krakow, Poland}
\address[e]{CERN, Geneva 23, Switzerland}

\date{ \today / Revised version: 1}

\begin{abstract}

We present the first estimates of isothermal compressibility (\kT) of
hadronic matter formed in relativistic nuclear collisions
($\sqrt{s_{\rm NN}} = 7.7$ GeV to
2.76~TeV) using experimentally observed quantities.
\kT~is related to the fluctuation in 
particle multiplicity, temperature, and volume of the
system formed in the collisions. 
Multiplicity fluctuations are obtained from the
event-by-event distributions of charged particle multiplicities in narrow
centrality bins. The 
dynamical components of the fluctuations are extracted
by removing the contributions to the fluctuations from the number of
participating nucleons. 
From the available experimental data, a constant
value of \kT~has been observed as a function of collision energy.
The results are compared with calculations from UrQMD, AMPT, and EPOS 
event generators, and estimations of \kT~are made for Pb-Pb collisions at the 
CERN Large Hadron Collider. 
A hadron resonance gas (HRG) model has been used to calculate \kT~as a
function of collision energy. Our results show a decrease in
\kT~at low collision energies to \sNN~$\sim$~20~GeV, beyond which
the \kT~values remain almost constant.
 
\end{abstract}



\end{frontmatter}

\section{Introduction}

The determination of the thermodynamic state of matter formed in
high-energy nuclear collisions is of
great importance in understanding the behaviour of the matter formed at high
temperature and/or energy density. 
A set of basic macroscopic quantities,
such as temperature, pressure, volume, entropy, and energy density, as
well as a set of response functions, including specific heat,
compressibility and different susceptibilities define the
thermodynamic properties of the system. These quantities are related
by the equation of state (EOS), which on the other hand, governs the
evolution of the system. One of the basic goals of calculating the
thermodynamic quantities, such as the 
specific heat ($c_v$) and isothermal compressibility (\kT) 
is to obtain the EOS of the matter~\cite{mrow,mekjian,stokic,wang,muller,lacey,sierk}.
The $c_v$ is the amount of energy per unit change in 
temperature and is related to the fluctuation in the temperature of 
the system~\cite{stodolsky,shuryak1}. 
The \kT~describes the relative variation of the volume of a system 
due to a change in the pressure at constant temperature. Thus \kT~is 
linked to density fluctuations and can be expressed in terms of 
the second derivative of the free energy with respect to the pressure. 
In a second order phase transition \kT~is expected to show a 
singularity. The determination of \kT~as well as $c_v$ can elucidate the 
existence of a phase transition and its nature.

Heavy-ion collisions at 
ultra-relativistic energies produce matter at extreme conditions of 
energy density and temperature, where a phase 
transition from normal hadronic matter to a deconfined state of 
quark-gluon plasma (QGP) takes place. 
Lattice QCD calculations have affirmed a crossover transition at zero 
baryonic chemical potential (\muB) ~\cite{aoki,baza}. On the other hand, 
QCD inspired phenomenological models~\cite{gottlieb,fukugita,schaefer,herpay} 
predict a first order phase transition at high \muB. 
This suggests the possible existence of a QCD critical point where the first order 
transition terminates. The current focus of theoretical and 
experimental programs is to understand the nature of the phase transition 
and to locate the critical point by exploring multiple signatures. 
Since \kT~is sensitive to the phase transition, its dependence on the 
\muB~or the collision energy provides one of the basic measurements on 
this subject.

Recently, collision energy dependence of $c_v$
has been reported by analysing the event-by-event mean transverse momentum
($\langle p_T\rangle$) distributions~\cite{sumit}. 
In this approach, the $\langle p_T\rangle$ distributions in finite \pT~ranges are
converted to distributions of effective temperatures. The
dynamical fluctuations in temperature are extracted by subtracting
widths of the corresponding mixed event distributions. 

In the present work, we have calculated the isothermal compressibility
of matter formed in high energy collisions using experimentally
observed quantities, as prescribed in Ref.~\cite{mrow}. 
This method uses the fluctuations of particle multiplicities produced in the 
central rapidity region. It may be noted that enhanced fluctuation of
particle multiplicity had 
earlier been proposed as signatures of 
phase transition and critical point~\cite{stephanov,heiselberg,gazdzicki,begun,mrow2}. 
Thus the study of event-by-event multiplicity fluctuations and 
estimation of \kT~are important for understanding the nature of matter at extreme 
conditions.
The experimental data of event-by-event multiplicity fluctuations at
the Relativistic Heavy-Ion Collider (RHIC) at Brookhaven
National Laboratory (BNL) and Super Protron Synchrotron (SPS) of CERN
have been used in combination with 
temperatures and volumes of the system at the chemical freeze-out to
extract the values of \kT.
These results are compared to that of three event generators and 
the hadron resonance gas (HRG) model. Our results 
provide important measures for the beam energy scan program of RHIC
and the experiments at the CERN Large Hadron Collider (LHC), and
gives guidance for experiments at the Facility for 
Antiproton and Ion Research (FAIR) at GSI and the Nuclotron-based Ion
Collider facility (NICA) at JINR, Dubna.

\section{Methodology}

Isothermal compressibility is the measure of
the relative change in volume with respect to change in
pressure~\cite{mrow},
\beqa
\left.k_T\right|_{T,\la N\ra} &=&
-\frac{1}{V}\left.\l\frac{\partial V}{\partial P}\r\right|_{T,\la N\ra}
\label{eq.kT}
\eeqa
where $V, T, P$ represent volume, temperature, and pressure of the
system, respectively, and $\la N\ra$ stands for the mean yield of the particles. 
In the Grand Canonical Ensemble (GCE) framework, the variance ($\sigma^{\rm 2}$)
of the number of particles ($N$)
is directly related to isothermal compressibility~\cite{mrow,adare}, i.e,
\beqa
\sigma^{\rm 2} = \frac{k_{\rm B}T \la N\ra ^{\rm 2}}{V}k_{\rm T},
\label{ket}
\eeqa
where $k_{\rm B}$ is the Boltzmann constant.
Charged particle multiplicity fluctuations have been characterised by the scaled variances of 
the multiplicity distributions, defined as,
\beqa
\omega_{\rm ch}=\frac{\langle N_{\rm ch}^{\rm 2} \rangle - \langle
  N_{\rm ch} \rangle^{\rm 2}}{\langle N_{\rm ch} \rangle}=\frac{\sigma^{\rm 2}}{\mu}
\eeqa
where $N_{\rm ch}$  is the charged particle multiplicity per event,
and $\mu = \la N_{\rm ch}\ra$.
Following the above two equations, we obtain, 
\beqa
\omega_{\rm ch} = \frac{k_{\rm B}T\mu}{V}k_{\rm T},
\label{imp}
\eeqa
which makes a connection between multiplicity fluctuation and \kT.
This formalism, using GCE properties, may be applied to experimental measurements at 
mid-rapidity, as energy and conserved quantum numbers are exchanged 
with the rest of the system~\cite{jeon}. 
At the chemical freeze-out surface, the inelastic collisions cease,
and thus the hadron multiplicities get frozen.  While the ensemble average
thermodynamic properties like the temperature and volume can be
extracted from the mean hadron yields, \kT~can be accessed through
the measurements of the event-by-event multiplicity fluctuations.

\section{Multiplicity fluctuations: experimental data}

\begin{figure}
\includegraphics[width=0.45\textwidth]{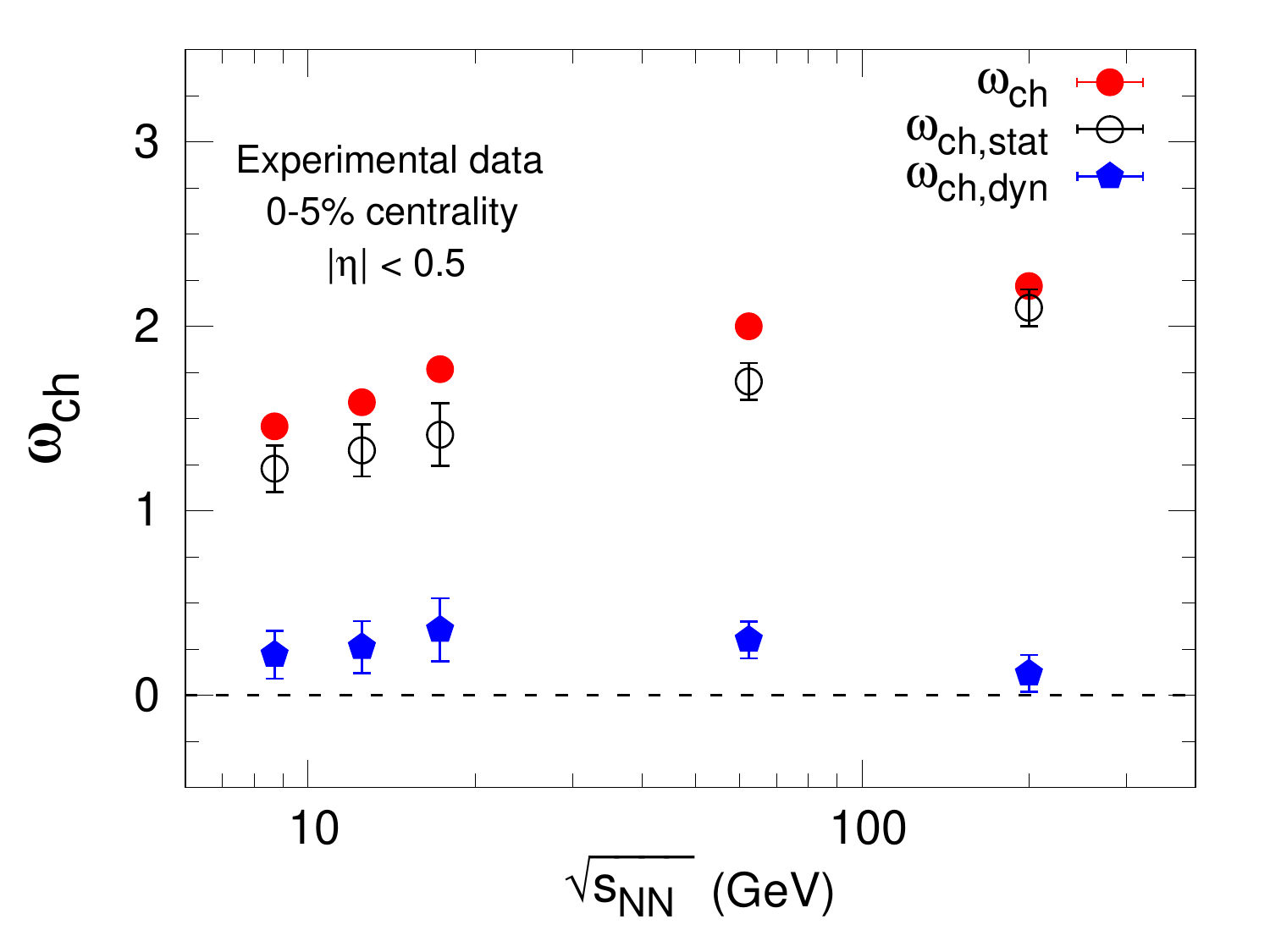}
\caption{Beam-energy dependence of scaled variances of multiplicity 
  distributions ($\omega_{\rm ch}$) for central (0-5\%) Au-Au (Pb-Pb) collisions from the 
  available experimental data~\cite{adare,abbott,alt,alt1,na61a,na61b,sako}. 
  The statistical components of fluctuations ($\omega_{\rm ch,stat}$) 
  using the participant model calculations have been shown. 
  The dynamical components of the fluctuations ($\omega_{\rm ch,dyn}$)
  are  obtained by subtracting 
  the statistical components from the measured values. 
}
\label{wch}
\end{figure}

The multiplicity fluctuations have been measured for a range of 
collision energies by the E802 collaboration~\cite{abbott} at BNL-AGS,
WA98~\cite{aggarwal}, NA49~\cite{alt,alt1}, NA61~\cite{na61a,na61b} and CERES~\cite{sako}
experiments at CERN-SPS, and PHENIX experiment~\cite{adare} at RHIC.
The results of these measurements could not be compared directly
because of differences in the kinematic acceptances and detection efficiencies.
The experimental results are normally reported after correcting for
detector efficiencies. But the acceptances in pseudorapidity~($\eta$)
need not be the same for these experiments. 
The results from the experiments have been scaled to 
mid-rapidity so that these can be presented in the same footing~\cite{adare, mait}.
Fig.~\ref{wch} shows the values of $\omega_{\rm ch}$ for
$|\eta|<0.5$ in central (0-5\%) collisions as a function of the
collision energy~\cite{mait}.
The solid circles represent experimental measurements. An 
increase in the scaled variances with the increase in collision energy
has been observed from these data.


It is to be noted that the widths of the charged particle distributions
and $\omega_{\rm ch}$ get their contributions from several sources,
some of which are of statistical in nature and the rest have dynamical origins.
The dynamical components are connected to thermodynamics and have been
used in the present work to extract \kT~\cite{mrow}. Thus an estimation of the statistical part is necessary to
infer about the dynamical component of multiplicity fluctuations.

One of the major contributions to statistical fluctuations comes from the
geometry of the collision, which includes variations in the impact
parameter or the number of participating nucleons. In a participant 
model~\cite{heiselberg}, the nucleus-nucleus collisions
are treated as superposition of nucleon-nucleon interactions. Thus the
fluctuation in multiplicity arises because of the fluctuation in
number of participants ($N_{\rm part}$) and the fluctuation in the
number of particles produced per participant. In this formalism, based
on Glauber type of initial conditions, $\omega_{\rm ch}$ can be expressed as, 
\beqa
\omega_{\rm ch} = \omega_{\rm n} + \langle n \rangle \omega_{N_{\rm part}},
\label{om}
\eeqa
where $n$ is the number of charged particles produced per participant, 
$\omega_{\rm n}$ denotes fluctuations in $n$, and $\omega_{N_{\rm part}}$ is  the
fluctuation in $N_{\rm part}$. 
The value of $\omega_{\rm n}$ has a
strong dependence on acceptance. The fluctuations in the number of
accepted particles ($n$) out of the total number of produced particles
($m$) can be calculated by assuming that the distribution of $n$ follows a
binomial distribution. This is given as~\cite{heiselberg,aggarwal},
\beqa
\omega_{\rm n} = 1 - f + f \omega_{\rm m},
\eeqa
where $f$ is the fraction of accepted particles. 
The values of $f$ and $\omega_{\rm  m}$ are obtained from
proton-proton collision data of the number of charged particles 
within the mid-rapidity range and the total number of charged
particles produced in the collision~\cite{whitmore,phobos,alice1,cms,aggarwal} . 
Using these, we obtain the values of 
$\omega_{\rm  n}$ as a function of collision energy. 

The values of $\omega_{\rm n}$ vary within 0.98 to 2.0
corresponding to \sNN~=7.7~GeV to 2.76~TeV, and are in agreement with
those reported for SPS energies~\cite{aggarwal}.
The distribution of $N_{\rm part}$ for narrow centrality bins yields
the value of $\omega_{N_{\rm part}}$. With the choice of narrow
bins in centrality selection, $\omega_{N_{\rm part}}$ values
remain close to unity from peripheral to central collisions.
With the knowledge of $\omega_{\rm n}$, $\langle n \rangle$ and
$\omega_{N_{\rm part}}$, the statistical components of 
$\omega_{\rm ch}$ from the participant model have been extracted. The
values of $\omega_{\rm ch,stat}$ are presented as open symbols in Fig.~\ref{wch} as a function of collision energy. 
The uncertainties  in $\omega_{\rm ch,stat}$ are derived from the
statistical and systematic uncertainties in $n$ and $\omega_{\rm  n}$.

The dynamical fluctuations of $\omega_{\rm ch}$ (denoted as $\omega_{ch,dyn}$)
are extracted by subtracting the statistical fluctuations from the
measured ones. In Fig.~\ref{wch}, the values of $\omega_{ch,dyn}$ are plotted (as diamond
symbols) as a function of collision energy.
Within the quoted errors, $\omega_{ch,dyn}$ is 
seen to remain constant as a function of collision energy. However,
a decreasing trend may be seen for $\sqrt{s_{\rm NN}} > 20$ GeV.
More experimental data at low and intermediate collision energies are needed to
conclude the nature of the fluctuations as a function of the collision
energy.

\section{Multiplicity fluctuations from event generators}

In order to validate the results from experimental data, we have
analysed the results from three different event generators, which are:
AMPT (A Multi Phase
Transport)~\cite{ampt,ampt2,ampt3}, UrQMD (Ultra-relativistic Quantum
Molecular Dynamics)~\cite{UrQMD1,UrQMD2}, and EPOS~\cite{EPOS1,EPOS3,EPOS4}. 
Multiplicity fluctuations using the AMPT model have been studied for
the default (DEF) and string melting (SM) modes~\cite{mait}.
In the default mode, hadronization takes place via the string fragmentation,
whereas in the SM mode, hadronization takes place via quark coalescence. 
The UrQMD is a
microscopic transport model, where the hadron-hadron interactions and the space-time
system evolution are studied based 
on the covariant propagation of all hadrons in combination with
stochastic binary scatterings, color string formation, and resonance
decay. UrQMD  has been previously used to simulate production of different particles
and analysis of their event-by-event
fluctuations~\cite{UrQMDapply1,UrQMDapply2,Bleichersus,Sahoo,Bhanu,Arghya}.

The EPOS(3+1) viscous hydrodynamical model incorporates multiple
scattering approach based upon the Gribov-Regge (GR) theory and
perturbative QCD~\cite{EPOS3}. The hydrodynamical
evolution starts from flux tube (or relativistic strings) initial conditions, generated  by the
GR framework. The string formation occurs due to initial
scatterings, which later breaks into segments identified as
hadrons. One of the salient features of the model is the
classification of two regions of physical interest on the basis of
density, such as core (high density) and corona (low density)~\cite{EPOS4}.  
For the centrality dependence of observables, the corona plays a major role at large rapidity
and low multiplicity events and contributes to hadronization. However,
for most central collisions, a core with collective hadronization is
created from corona because of a large number of nucleons suffering
inelastic collisions. Results from EPOS match experimental data at
RHIC and LHC for particle multiplicities, transverse momenta and 
correlation patterns~\cite{EPOS1,EPOS3,EPOS4,EPOS5}.

For the present study, a large number of events are generated using the
event generators for Au-Au collisions between \sNN = 7.7
to 200 GeV, corresponding to the RHIC energies, and
for for Pb-Pb collisions at \sNN~= 2.76~TeV.
In all cases, the centrality of the collision has been
selected using minimum bias distributions of charged particle multiplicities in the
range,  $0.5 < |\eta| < 1.0$.  The
multiplicities and multiplicity fluctuations have been obtained within
the kinematic range, $|\eta| < 0.5$ and $0.2 < p_{\rm T}<2.0$~GeV/c.  
The $\eta$-range used for the centrality selection is different from
the one for the fluctuation study, and thus poses almost no bias on the
fluctuation analysis.
To minimise the geometrical fluctuations, calculations are first done for
narrow (1\%) centrality bins. These results are then combined to
make wider bins by using centrality
bin width correction method which takes care of the impact
parameter variations~\cite{mait}.

\begin{figure}
\includegraphics[width=0.49\textwidth]{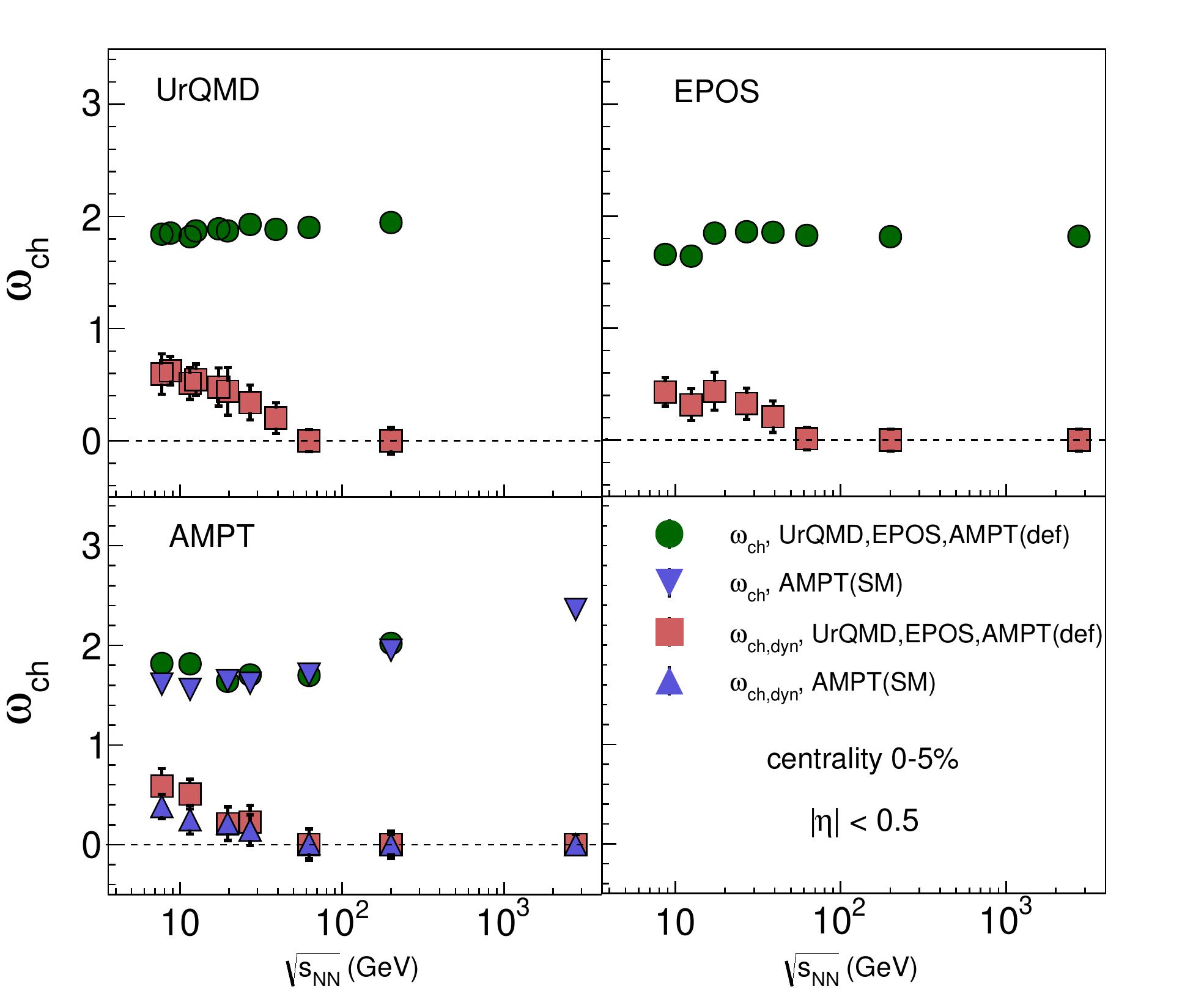}
\caption{
Collision energy dependence of scaled variances of charged particle multiplicity
distributions for central (0-5\%) Au-Au (Pb-Pb) collisions from event generators, 
UrQMD, EPOS and AMPT.  The dynamical multiplicity fluctuations
($\omega_{ch,dyn}$) are obtained after subtracting the statistical
fluctuations from participant model.
}
\label{wch_model}
\end{figure}

Fig.~\ref{wch_model} shows the collision energy dependence of
$\omega_{ch}$ for central (0-5\%) collisions from the
event generators. Statistical
errors are calculated using the Delta 
theorem~\cite{delta}. It is observed that the fluctuations remain
somewhat constant over the energy range considered, except for the AMPT
events, where a small rise is seen at higher energies. 
The statistical components of the fluctuations have been calculated
from the participant model calculations, using the same procedure as
discussed in the previous section.
The dynamical components, $\omega_{ch,dyn}$,  are obtained after
subtracting the statistical fluctuations, and are also shown in the Fig.~\ref{wch_model}.
In all cases, the dynamical
multiplicity fluctuations decrease with the increase of high collision
energy to $\sqrt{s_{\rm NN}} > 62.4$~GeV, beyond which the
fluctuations are close to zero.

\section{\kT~from HRG model}

The values of \kT~can be obtained by employing a hadron resonance gas
model, which is based on a list of
majority of the hadrons and their resonances as per the
Particle Data Book~\cite{PDG}. It works within the framework of a multiple species non-interacting 
ideal gas in complete thermal and chemical equilibrium~\cite{alba,andronic,cleymans}. 
The HRG model
has been found to provide a good description of the mean hadron
yields using a few thermodynamic parameters at freeze-out (for a
recent compilation of the freeze-out parameters, see Ref.~\cite{sandeep1}).
The goal in the HRG model calculation is to obtain
\kT~directly from eqn.~1, where instead of total number of charged particles,
the attempt has been to calculate in terms of species dependence ($i$)
of the hadrons. The differential for the pressure $P\l T, \{\mu_i\}\r$
can be written as,
\beqa
dP = \l\frac{\partial P}{\partial T}\r dT + \sum_i\l\frac{\partial
  P}{\partial \mu_i}\r d\mu_i\label{eq.diffP1}, 
\eeqa
and so: 
\beqa
\left.\l\frac{\partial P}{\partial V}\r\right|_{T,\{\la N_i\ra\}} = \sum_i\l\frac{\partial P}{\partial \mu_i}\r
\left.\l\frac{\partial \mu_i}{\partial V}\r\right|_{T,\{\la N_i\ra\}}\label{eq.diffP2}.
\eeqa
While the first factor is straightforward to compute from the expression for $P$, the 
second factor $\left.\l\frac{\partial \mu_i}{\partial V}\r\right|_{T,\{\la N_i\ra\}}$ is obtained 
from the condition of constancy of $N_i$ as follows,
\beqa
dN_i = \l\frac{\partial N_i}{\partial T}\r dT + \l\frac{\partial N_i}{\partial V}\r dV + 
\l\frac{\partial N_i}{\partial \mu_i}\r d\mu_i\label{eq.diffNi}.
\eeqa
For fixed $N_i$, the above equation becomes,
\beqa
\left.\l\frac{\partial \mu_i}{\partial V}\r\right|_{T,\{\la N_i\ra\}}  = -\frac{\l\frac{\partial N_i}{\partial V}\r}
{\l\frac{\partial N_i}{\partial \mu_i}\r}\label{eq.dmudv}.
\eeqa
Within HRG, $\frac{\partial N}{\partial V}=\frac{\partial P}{\partial\mu}$. Thus, Eq.~\ref{eq.diffP2} becomes
\beqa
\left.\l\frac{\partial P}{\partial V}\r\right|_{T,\{\la N_i\ra\}} 
= -\sum_i\frac{\l\frac{\partial P}{\partial \mu_i}\r^2}{\l\frac{\partial N_i}{\partial \mu_i}\r}
\eeqa
which is used to get $k_T$ using Eq.~\ref{eq.kT},

\beqa 
\left.k_T\right|_{T,\{\la N_i\ra\}} =
\frac{1}{V}\frac{1}{\sum_i{\l\frac{\partial P}{\partial \mu_i}\r^2}/{\l\frac{\partial N_i}{\partial \mu_i}\r}}. 
\eeqa

\begin{figure}
\includegraphics[width=0.49\textwidth]{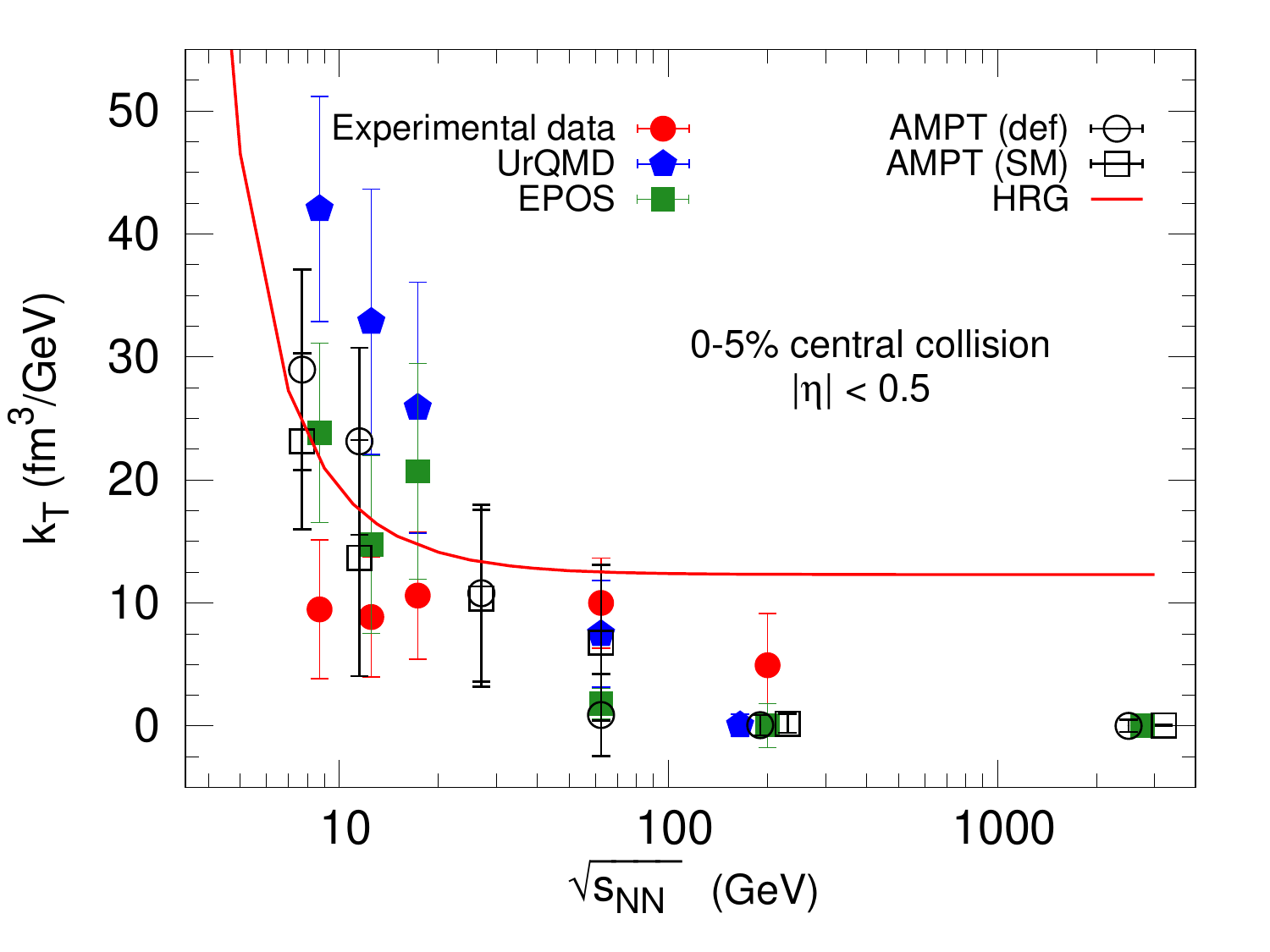}
\caption{Isothermal compressibility, \kT, as a function of \sNN~ for 
  available experimental data for central (0-5\%) Au-Au (Pb-Pb) collisions. 
  Results for three event generators are presented.
  Results from HRG calculations are superimposed. 
  }
\label{final_kt}
\end{figure}

This prescription of the HRG model has been used to calculate \kT~for
Au-Au collisions as a function of collision energy, which are
presented in terms of the solid curve in Fig.~\ref{final_kt}.
With the increase of collision energy, the values of \kT~decrease up to
\sNN~$=$20~GeV. 
However, at higher energies, \kT~remains almost constant.
This follows primarily from the behaviour of chemical freeze-out temperature as
a function of collision energy.

\section{Compilation of \kT}

Finally, the values of \kT~are calculated from the available
experimental data and event generators using 
the dynamical fluctuations, $\omega_{\rm ch,dyn}$, which are presented in the
figures~\ref{wch} and \ref{wch_model}. The mean charged particle multiplicities are
obtained under the same kinematic conditions.
The calculation of \kT~requires temperature and volume, which are
obtained from different sets of measurements. 
The chemical freeze-out temperature ($T_{\rm ch}$) and the corresponding volume of the system 
have been obtained by
fitting the measured identified particle yields using thermal model
calculations~\cite{cleymans,sandeep1,cleymans2,pbm,star1,alice2}.
For the calculation of \kT, both $T_{\rm ch}$ and $V$
have been obtained from Ref.~\cite{sandeep1}.

A compilation of \kT~as a function of \sNN~for central Au-Au (Pb-Pb)
collisions is presented in Fig.~\ref{final_kt}.  In the
absence of experimental data at the LHC, calculations from AMPT and EPOS  have 
been presented. 
From the available experimental data, it is observed that, $k_{\rm T}$~remains
almost constant within the assigned errors. The
results from the event generators are seen to decrease with an
increase in the
collision energy and remain constant at higher energies.
The results from HRG calculations show a sharp decrease in \kT~at low collision energies.
Thus more experimental data points at collision
energies below \sNN~$\sim$~20~GeV are needed to validate our findings.

The extraction of \kT~may be affected by several sources of
uncertainty. The evaluation of the statistical component of the
fluctuation poses one of the largest uncertainties. We have
used a participant model calculation to obtain the
$\omega_{ch,stat}$ based on the Glauber type of initial
conditions. Another effect which affects
the charged particle production is the resonance decay of particles.
This is studied for Au-Au collisions at \sNN~=~200 GeV
and Pb-Pb collisions at \sNN~=~2.76~TeV using AMPT and EPOS event
generators by turning off and on the higher order resonances.
The differences between the two cases are very small and within the errors, implying that 
resonance decay effects are negligible for multiplicity fluctuations.
Other sources of fluctuations which affect the extraction of
$\omega_{ch,dyn}$ include uncertainty in the initial state
fluctuations and fluctuations in the amount of stopping.
In view of the uncertainties from different sources 
which could not be considered presently, the extracted values are the upper limits of \kT.

\section{Summary}

We have studied the isothermal
compressibility of the system formed at the time of chemical
freeze-out in relativistic nuclear collisions for \sNN~from 7.7~GeV to 2.76~TeV.
We have shown that \kT~is related to the fluctuation in particle
multiplicity in the central rapidity region. Multiplicity fluctuations
have been obtained from available experimental data and event
generators. The dynamical fluctuations are extracted
from the total fluctuations
by subtracting the statistical components using 
contributions from the number of
participating nucleons. 
For the calculation of \kT, the temperature and volume were taken from
the thermal model fits of the measured particle yields
at the chemical freeze-out. 
Within quoted errors, the values of \kT~from the experimental
data remain almost constant as a function of energy. 
Using the event generators, we have seen that \kT~decreases with
an increase of the collision
energy. The estimation of \kT~presented in the present manuscript relies on several
assumptions, most importantly on the estimation of dynamical
fluctuations. The results of \kT~represent the upper limits because of
unknown contributions to the statistical components.

We have calculated the values of \kT~from the HRG model for a wide
range of collision energy.  With the increase of collision energy,
\kT~values decrease up to \sNN~$\sim$~20~GeV, beyond which
the \kT~remain almost constant.
The nature of \kT~as a function of collision energy is similar to what
has been observed for $c_v$~\cite{sumit}. 
A higher value of \kT~at low energies compared 
to higher energies indicates that the collision system is more
compressible at the lower energies.
This study gives a strong impetus for the second phase of the beam energy scan program of
RHIC and planned experiments at FAIR and NICA.



\medskip

\noindent{\bf Acknowledgements}\\

The authors would like to thank 
Stanislaw Mrowczynski, Jean Cleymans, Victor Begun and
Pradip K. Sahu for discussions on the concepts leading to this work. 
SPA is grateful to Klaus Werner for providing the EPOS code.
MM is thankful to the High Energy Physics group of Bose Institute for
useful discussions. SB wishes to thank Claude A. Pruneau for fruitful discussions 
 during the preparation of the manuscript. 
SB is supported by the U.S.Department of Energy Office of Science, 
Office of Nuclear Physics under Award Number DE-FG02-92ER-40713.
SC is supported by the Polish Ministry of Science and Higher 
Education (MNiSW) and the National Science Centre grant 2015/17/B/ST2/00101.
This research used resources of the LHC grid computing centers at
Variable Energy Cyclotron Center, Kolkata and at Bose Institute, Kolkata.

\medskip
\noindent{\bf References}\\

\end{document}